\documentclass[prb,preprint,preprintnumbers,amsmath,amssymb,showpacs]{revtex4}

\usepackage{graphicx,bm,epsfig}

\makeatletter
\def\graphicscale{\twocolumn@sw{0.33}{0.4}}
\makeatother

\def\spose#1{\hbox to 0pt{#1\hss}}
\def\lesssim{\mathrel{\spose{\lower 3pt\hbox{$\mathchar"218$}}
 \raise 2.0pt\hbox{$\mathchar"13C$}}}
\def\gtrsim{\mathrel{\spose{\lower 3pt\hbox{$\mathchar"218$}}
 \raise 2.0pt\hbox{$\mathchar"13E$}}}

\def\<{\langle}
\def\>{\rangle}

\begin{document}

\title{Predicting the thermodynamics by using state-dependent interactions}

\author{Giuseppe D'Adamo\footnote{Address after December 1, 2012:
Dip. di Fisica, Sapienza Universit\`a di Roma, I-00185 Roma, Italy.} }
\email{giuseppe.dadamo@aquila.infn.it}
\affiliation{Dipartimento di Scienze Fisiche e Chimiche, Universit\`a dell'Aquila,
V. Vetoio 10, Loc. Coppito, I-67100 L'Aquila, Italy}
\author{Andrea Pelissetto}
\email{andrea.pelissetto@roma1.infn.it}
\affiliation{Dipartimento di Fisica, Sapienza Universit\`a di Roma and
INFN, Sezione di Roma I, P.le Aldo Moro 2, I-00185 Roma, Italy}
\author{Carlo Pierleoni}
\email{carlo.pierleoni@aquila.infn.it}
\affiliation{$^3$ Dipartimento di Scienze Fisiche e Chimiche, Universit\`a dell'Aquila and
CNISM, UdR dell'Aquila, V. Vetoio 10, Loc. Coppito, I-67100  L'Aquila, Italy}

\begin{abstract}
We reconsider the structure-based route to coarse graining in which
the coarse-grained model is defined in such a way to reproduce 
some distributions functions of the original system as accurately as 
possible.
We consider standard expressions for pressure and chemical potential
applied to this family of coarse-grained models with density-dependent 
interactions and show that they only provide approximations to the 
pressure and chemical potential of the underlying original system.
These approximations are then carefully compared in two cases: 
we consider a generic microscopic system in the low-density regime
and polymer solutions under good-solvent conditions.
Moreover, we show that the state-dependent
potentials depend on the ensemble in which they have been derived. Therefore,
care must be used in applying canonical state-dependent potentials to predict
phase lines, which is typically performed in other ensembles.
\end{abstract}


\maketitle

\section{Introduction}

In condensed matter physics, chemistry, and material science 
state-dependent interactions arise in many different 
contexts. 
They are usually obtained from coarse-graining procedures in which
a subset of the degrees of freedom is integrated out
(see Refs. \onlinecite{Voth2009, PCCP2009, SM2009, FarDisc2010} 
for an overview of methods and applications). 
Indeed, any coarse graining of 
the original (be it classical or quantum) system induces many-body interactions
among the remaining degrees of freedom. The idea is then to replace these 
complex interactions  with state-dependent potentials that are 
usually taken to be 
pairwise additive for computational efficiency
(if necessary, three-body forces can also be introduced, as in 
Ref.~\onlinecite{LLV-10})
and that are therefore 
much more tractable from a theoretical
and/or numerical point of view. 
Different criteria have been used to select the optimal 
set of state-dependent pair potentials. In the structural route the 
model with state-dependent interactions is required to reproduce 
some distribution functions associated with the coarse-grained (CG)
degrees of freedom.
\cite{RS-67,Rowlinson-67,Barker,Casanova,VanderHoef,%
LBHM-00,BL-02,MullerPlathe-02,PK-09}
It is also possible to define the CG model by matching
the forces \cite{force-matching1,force-matching2,force-matching3} 
acting on the CG sites
computed in the original, underlying system, 
or by requiring the CG model to reproduce solvation free energies. 
Also state-dependent potentials suitable to treat 
inhomogeneous systems have been 
proposed.\cite{Pagonabarraga,Rutledge} We should further mention 
mixed coarse-graining strategies that try
to match simultaneously the pair distribution function and some other 
thermodynamic property, for instance, by constraining the virial pressure 
to be equal to the pressure of the microscopic model.
\cite{Voth2009, PCCP2009, SM2009, FarDisc2010, Rutledge} 
These mixed approaches, although in principle incorrect
since the potential is uniquely
defined by the pair distribution function $g({\bf r})$ and 
the density according to 
Henderson's theorem,\cite{Henderson-74} may still be of value in practical 
numerical calculations. Indeed, $g({\bf r})$ is only known 
with statistical errors
and is little sensitive to the tail of the potential:
visibly different potentials may produce
structures with essentially indistinguishable pair distribution functions. 
\cite{MullerPlathe-02}
Therefore, the large-distance behavior
of the CG interactions is determined with a large uncertainty, 
which might leave 
some flexibility to implement an additional constraint. We will not
discuss these approaches any further, focusing on the 
conceptual problems of the approach rather than on the difficulties of 
practical numerical implementations.

In this paper we discuss the structural approach, which dates back to the
early days of liquid state theory, 
\cite{RS-67,Rowlinson-67,Barker,Casanova,VanderHoef}
considering state-dependent pair interactions 
for  a very general class of classical systems. The microscopic,
underlying model 
to which the coarse-graining procedure is applied consists 
of polyatomic molecules. No hypothesis is made on the nature of 
the interactions among the atoms: we only require them
to be state independent, but, otherwise, they are arbitrary and can, 
in principle, include any type of many-body terms. Our discussion
therefore applies both to monoatomic systems with three- and higher-body
interactions (for instance, to noble gases whose accurate study requires the 
introduction of the three-body Axilrod-Teller potential
\cite{AT-43}) and to soft-matter systems, such as polymers, in which 
the complexity lies in the number of atoms involved rather than in the 
many-body nature of the atom-atom interactions.
\cite{LBHM-00,Likos-01,PK-09} Within the CG approach
we replace the microscopic system at a given thermodynamic state with a
CG system of monoatomic molecules that interact by means 
of state-dependent pair potentials. The latter are fixed by 
a structural requirement, the equality of a specific pair distribution 
function at a given thermodynamic state. 
If the underlying system is formed by monoatomic molecules, 
we consider the usual radial pair distribution function. In the 
case of polyatomic systems, 
each molecule is replaced by a CG monoatomic molecule located at some 
point $X$. In many instances, $X$ corresponds to the center of mass 
of the original molecule, but other choices are possible: for instance, in CG 
models for star polymers the point $X$ usually 
coincides with the center of the star.
\cite{LLWAJAR-98,WLL-99,JDLFL-01,DLL-02,Likos-01,DAdamo2012}
For our purposes, we do not need to specify how $X$ is chosen. 
We only require the coordinates ${\bf r}_X$ of $X$ to be a weighted 
average of the 
positions of the atoms belonging to the molecule. Once $X$ is chosen,
we can consider the $X$-$X$ pair distribution function 
\begin{eqnarray} 
   g({\bf r}_1 - {\bf r}_2) = 
   \left\langle {1\over N\rho^2} \sum_{ij} \delta({\bf r}_{X,i} - {\bf r}_1) 
                      \delta({\bf r}_{X,j} - {\bf r}_2) \right\rangle,
\label{defg}
\end{eqnarray}
where $\rho$ is the density and $N$ the number of molecules. 
The state-dependent potential 
is fixed by the requirement that the pair distribution 
function $g_{CG}({\bf r})$ in the 
CG model is equal to $g({\bf r})$ in the underlying system at the same 
density.
From a practical point of view, the determination of the potential 
is not an easy task and several method have been devised, 
like the iterative Boltzmann inversion method\cite{RPMP-03} and
the inverse Monte Carlo method.\cite{LL-95,Soper-96} 
Variational approaches have also been discussed, optimizing the 
coefficients of suitable parametrizations of the CG potentials.
\cite{MBFRM-00,AB-01}
Note that the structural approach we have described here 
is not the only one that 
is used in practical applications. We should also mention 
the force-matching approach (often called multiscale coarse-graining
method \cite{force-matching1}), in which
the state-dependent potential is determined 
by requiring the CG system to match the atomistic force on the CG atoms as 
accurately as possible. Also this method has a structural 
interpretation: the matching condition is equivalent\cite{force-matching3} 
to require 
the CG force to satisfy the appropriate Yvon-Born-Green equation
\cite{HansenMcDonald} that 
relates the pair and the three-body correlation function.
We mention that general theoretical formalisms within which 
different CG approaches can be derived are presented in 
Refs.~\onlinecite{Shell-08,KL-11}.
For a summary of methods and a detailed comparison in 
a specific example, see Ref.~\onlinecite{RJLKA-09}. 

Once a CG model has been defined, one may use it to predict 
thermodynamic (and, in the case of more complex CG systems, also
structural) properties for the underlying model. For instance, one would 
like to obtain estimates of the pressure of the microscopic system, by
studying (for example, by performing numerical simulations) the simpler
CG model.

Structurally derived
state-dependent potentials have been mostly discussed in the 
context of the canonical ensemble, see
Refs.~\onlinecite{SST-02,Louis-02,JHGL-07,Tejero} 
and references therein.
In this case, the potentials become 
density and temperature dependent. For this reason,
some thermodynamic relations, which can be rigorously proved 
for systems with state-independent interactions,
are no longer satisfied: in particular, 
the compressibility route and the virial route to the pressure are 
no longer equivalent for the CG system. 
This gave rise to several investigations concerning 
the thermodynamic consistency of models with state-dependent potentials.
\cite{SST-02,Louis-02,Tejero}
In this paper we again investigate the conceptual issues which arise 
when considering state-dependent interactions.
The first question 
we wish to address is whether the knowledge of the CG model 
at density $\rho$ 
(i.e., of the model with potentials obtained by matching the pair 
distribution function at density $\rho$) allows one to obtain 
informations on the thermodynamical behavior of 
the underlying system at the {\em same} value of $\rho$.
Considering, for instance, pressure and chemical potential, the idea 
is to compute these two quantities in the CG model. The results of this
calculation are 
then taken as estimates of the pressure and chemical potential of the 
underlying system. The main problem in implementing this strategy stems 
from the fact that there is no unambiguous way to define thermodynamic
quantities
in the presence of state-dependent potentials: Approaches that are equivalent
for state-independent interactions give different espressions for 
pressure and chemical potential
if potentials depend on the thermodynamic state. Following 
Ref.~\onlinecite{SST-02}, we shall consider two different approaches. 
In the passive approach, the pressure is estimated by using the usual
virial expansion, while in the active approach one derives the 
generalized expression of Ascarelli and Harrison.\cite{AH-69} 
Both approaches are not  thermodynamically consistent and only 
provide an approximation of the pressure of the underlying model.
A quantitative comparison shows that 
the passive approach provides the most accurate estimates, 
i.e., it best reproduces the pressure of the underlying model.
The same result
holds for the chemical potential. Widom's method applied to the CG system, 
even though it does not reproduce the underlying-model value, provides an 
estimate of the chemical potential that is closest to the underlying-model 
value than the estimate obtained in the active approach.
Note that, if we assume that the 
density-dependent (DD) potentials are known for {\em all} densities 
$0\le \rho \le \rho_{\rm max}$ and no phase transitions occur in
such density interval, then, via the compressibility route, 
we can compute the exact equation of state of the underlying model for 
any $\rho\le \rho_{\rm max}$, hence derive the exact pressure, 
chemical potential, and 
free energy. However, in this case CG models with DD
potentials play little role. Indeed, their computation requires a study of the 
microscopic  model in the whole density interval. 
But this study would also provide the equation of state 
directly, without the necessity of introducing any sort of CG model.

It is interesting to extend the analysis of the thermodynamic properties 
of state-dependent potentials to other ensembles.
For instance, for phase-coexistence studies it is more natural to consider
the grand-canonical ensemble, while the isothermal-isobaric ensemble may be the 
best suited to interpret experimental data, since pressure is fixed 
in experiments. In these ensembles one would consider fugacity- and pressure-
dependent pair potentials, respectively. 
Again, the question we wish to address is how 
to obtain thermodynamic predictions in these different ensembles by using 
state-dependent interactions.

The paper is organized as follows. In Sec.~\ref{sec2} we discuss
density-dependent potentials and the consistency of several commonly
used methods to determine pressure and chemical potential. We show 
that none 
of them is exact: none of the thermodynamic quantities extracted via a CG 
approach is more than an approximate estimate. 
In Sec.~\ref{sec3} we study the accuracy of the different approximations:
first, in Sec.~\ref{sec3.1} we present a general, model-independent
discussion in the low-density limit, then, in Sec.~\ref{sec3.2} 
we present a specific calculation for linear polymers under good-solvent
conditions. The analysis of Sec.~\ref{sec2} is extended to the 
grand-canonical ensemble in Sec.~\ref{sec4}, while in Sec.~\ref{sec5}
we discuss the accuracy with which the thermodynamic behavior 
is reproduced by using fugacity-dependent potentials. 
Finally, in Sec.~\ref{sec6} we present our conclusions.

\section{Density-dependent potentials in the canonical ensemble} \label{sec2}

We start by considering the canonical ensemble and discuss density-dependent
interactions. In principle, one should also consider a temperature dependence.
However, temperature does not play a role in the discussion (we do not discuss
temperature derivatives of the free energy), so that it will be omitted for simplicity.
In the structural approach to coarse graining,
the basic quantity of interest is the $X$-$X$ pair distribution
function $g({\bf r};\rho)$ defined in Eq.~(\ref{defg}) for the underlying model.
Knowledge of $g({\bf r};\rho)$ allows us to 
obtain the compressibility of the system by using the compressibility relation,
\cite{HansenMcDonald} which holds for any choice of the CG site $X$:
\begin{equation}
{1\over K(\rho)} = 1 + \rho \int d^3{\bf r}\, [g({\bf r};\rho) - 1], 
\label{compr_rel}
\end{equation}
where 
\begin{equation}
   K(\rho) = {\partial \beta P\over \partial \rho}
\end{equation}
is related to the isothermal compressibility $\chi_T = \beta/(\rho K)$ 
and $\beta = 1/k_B T$. No other thermodynamic quantity at density $\rho$ 
can be computed 
directly from $g({\bf r};\rho)$ or $K(\rho)$. However, they can be determined
if $K(\rho)$ is known along a thermodynamic path that starts at a state 
point at which the Helmholtz free energy is known. For instance,
in the absence of phase transitions in the density interval $[0,\rho]$,
other thermodynamic quantities like pressure, chemical potential, and 
Helmholtz free energy at density $\rho$ can then be obtained from $K(\rho)$ as 
\begin{eqnarray}
\beta P(\rho) &=& \int_0^\rho d\sigma\, K(\sigma), 
   \label{pFA-KFA} \\
\beta \mu(\rho) &=& \ln {\rho\over q_1} + 
     \int_0^\rho {d\sigma\over \sigma} (K(\sigma) - 1),
\label{muFA-KFA} \\
{\beta F\over N} = \beta f(\rho) &=& 
 \beta \mu(\rho) - {\beta P(\rho)\over \rho},
\label{fFA-KFA} 
\end{eqnarray}
where $q_1 = Z_1/V$ and $Z_1$ is the partition function of a single,
isolated molecule.
For convenience, here and in the following 
we set the de Broglie thermal length equal to one.

In the CG approach we discuss here, 
one maps each molecule onto a point particle.
Under the pair potential approximation, the CG molecules interact by means 
of the DD potential $V_{D,CG}({\bf r};\rho)$,
which is defined such as to reproduce the finite-density pair distribution
function $g({\bf r};\rho)$. 
Once $V_{D,CG}({\bf r};\rho)$ has been determined, one can use the 
CG model to compute thermodynamic quantities at the same density $\rho$. 
In principle one could also use it to study the thermodynamic behavior 
at densities $\rho'\not=\rho$, i.e., one could {\em transfer} the potentials
from one density to the other. The accuracy of this procedure, i.e.,
the transferability of the potentials, is an important question for 
CG studies, which, however, will not be considered here. 

In order to compute thermodynamic quantities in the CG model, 
we will consider two different approaches, which, following
Ref.~\onlinecite{SST-02}, will be referred to as the passive and the active
approach. 
To define the two approaches, we consider the more general
CG system at {\em fixed} interaction potential in the canonical
ensemble, i.e. at fixed $V$ and $T$. The Helmholtz free energy 
is given by
\begin{eqnarray}
&&F_{CG}(N,V,T;\rho_p) = N f_{CG}(\rho;\rho_p) 
\label{FCG-def}
\\
&& \quad = - k_B T\ln {1\over N!} \int d{\bf r}_1\ldots d{\bf r}_N \, 
   e^{-\beta\sum_{ij} V_{D,CG}({\bf r}_i - {\bf r}_j;\rho_p)},
 \nonumber 
\end{eqnarray}
where $\rho = N/V$. The free energy (\ref{FCG-def}) 
depends on two densities: $\rho = N/V$ 
is the usual quantity, while $\rho_p$ is the density parametrizing the pair
potential. No relation between $\rho$ and $\rho_p$ is assumed at this stage.
Using Eq.~(\ref{FCG-def}) we start by defining the pressure and the 
chemical potential for the general CG model at fixed $\rho_p$:
\begin{eqnarray}
P_{CG}(\rho,\rho_p) = 
- \left( {\partial F_{CG}\over \partial V} \right)_{N,T,\rho_p}, 
\\
\mu_{CG}(\rho,\rho_p) = 
\left( {\partial F_{CG}\over \partial N} \right)_{V,T,\rho_p}. 
\end{eqnarray}
If we define 
\begin{equation}
K_{CG}(\rho,\rho_p) = 
\left( {\partial\beta P_{CG}\over \partial \rho}\right)_{T,\rho_p},
\end{equation}
we obtain the standard thermodynamic relations
\begin{eqnarray}
\beta \mu_{CG}(\rho,\rho_p) &=& \ln \rho + 
   \int_0^\rho {d\sigma\over \sigma} (K_{CG}(\sigma,\rho_p) - 1), 
\label{muCG-K}
\\
\beta P_{CG}(\rho,\rho_p) &=& \int_0^\rho d\sigma\, 
     K_{CG}(\sigma,\rho_p),
\label{pCG-K}
\\
\beta f_{CG} (\rho,\rho_p) &=& 
   \beta \mu_{CG}(\rho,\rho_p) - {\beta P_{CG}(\rho,\rho_p)\over \rho}.
\label{fCG-K}
\end{eqnarray}
These expressions are the analog of those appearing in 
Eqs.~(\ref{pFA-KFA}), (\ref{muFA-KFA}), and 
(\ref{fFA-KFA}), with $q_1 = 1$, 
since the CG model we consider consists of monoatomic molecules.
Since $\rho_p$ is a fixed parameter, 
the pressure $P_{CG}(\rho,\rho_p)$ and the chemical potential 
$\mu_{CG}(\rho,\rho_p)$ can be determined
by using standard methods. 
In particular, the usual relation between pressure and virial holds. Hence
we have \cite{HansenMcDonald}
\begin{eqnarray}
&& \!\!\!\! \beta P_{CG}(\rho,\rho_p) = 
\beta P_{\rm vir}(\rho,\rho_p) = 
    \nonumber \\
&& \quad \rho - {2\pi\beta\rho^2\over3} 
   \int_0^\infty {\partial V_{D,CG}({\bf r};\rho_p)\over \partial r} 
     g({\bf r};\rho,\rho_p)\, r^3 dr.
\label{Virialpressure}
\end{eqnarray}
As for the chemical potential, since $\rho_p$ is fixed,
$\mu_{CG}(\rho,\rho_p)$ 
can be obtained by using 
Widom's insertion method\cite{Widom-63}
\begin{eqnarray}
&& \!\!\!\! \beta \mu_{CG}(\rho,\rho_p) = 
   \ln \rho - \nonumber \\
&& \quad \ln \left[ {1\over V} 
     \int d^3{\bf r}_{N+1}\, 
     \langle e^{-\beta U_{N+1}({\bf r}_{N+1};\rho_p)} \rangle_{N,V}
     \right], 
\end{eqnarray}
where $\langle\cdot\rangle_{N,V}$ is the canonical ensemble average
over $N$ molecules in a volume $V$, $\rho=N/V$, and 
$U_{N+1}({\bf r}_{N+1};\rho_p)$ is 
the insertion energy of an additional molecule 
computed by using $V_{D,CG}({\bf r};\rho_p)$.

The DD pair potential at density $\rho_p = \rho$ 
is determined by requiring the pair distribution
function $g_{CG}({\bf r};\rho,\rho)$ to be equal to the 
$XX$ pair distribution function $g({\bf r;\rho})$ in the underlying 
microscopic model:
$g_{CG}({\bf r};\rho,\rho) = g({\bf r;\rho})$. 
In the passive approach\cite{SST-02} $\rho_p$ is considered as a fixed parameter
which is set equal to $\rho$ only at the end of the calculations. 
Therefore, the pressure for the CG model 
at density $\rho$ is simply $P_{CG}(\rho,\rho)$;
analogously the chemical potential is defined as $\mu_{CG}(\rho,\rho)$.
Because of the compressibility
relation (\ref{compr_rel}), the CG and the microscopic 
compressibilities are equal, i.e.
\begin{equation}
     K(\rho) = K_{CG}(\rho,\rho).
\label{K-equality}
\end{equation}
This equality does not extend, however, to the other thermodynamic quantities:
both $\mu_{CG}(\rho,\rho)$ and $P_{CG}(\rho,\rho)$ differ from the 
chemical potential and pressure at density $\rho$ 
of the microscopic model. Indeed, 
barring unexpected concidences, it is expected in general
\begin{eqnarray}
&& \beta \mu^{\rm exc}(\rho) = 
\int_0^\rho {d\sigma\over \sigma}\, (K_{CG}(\sigma,\sigma)-1) \nonumber \\
&& \qquad \not= 
\beta \mu^{\rm exc}_{CG}(\rho,\rho) = 
\int_0^\rho {d\sigma\over\sigma}\, (K_{CG}(\sigma,\rho) - 1), \\
&& \beta P(\rho) = 
\int_0^\rho {d\sigma}\, K_{CG}(\sigma,\sigma) \nonumber \\
&& \qquad \not= 
\beta P_{CG}(\rho,\rho) = 
\int_0^\rho {d\sigma}\, K_{CG}(\sigma,\rho), 
\label{PCG-KCG}
\end{eqnarray}
where we used Eq.~(\ref{K-equality}) to replace $K(\rho)$ with
$K_{CG}(\rho,\rho)$ in the left-hand side integrals,
which give chemical potential and pressure in the underlying model.
Hence, if we use the virial route for the pressure or Widom's
insertion method for the chemical potential in the CG model, 
we only obtain 
approximations to the underlying-model pressure and 
chemical potential: even with the use of DD potentials, we cannot obtain the 
underlying-model thermodynamics at density $\rho$ from the study of the 
CG model at that density. Moreover, the approach is not thermodynamically
consistent. Indeed, Eq.~(\ref{PCG-KCG}) implies immediately that the 
density derivative of $P_{CG}(\rho,\rho)$ differs from $K_{CG}(\rho,\rho)$.
Of course, the pressure and chemical
potential of the underlying system 
can be obtained by using $K(\rho) = K_{CG}(\rho,\rho)$ and relations 
(\ref{pFA-KFA}) and (\ref{muFA-KFA}).\cite{Louis-02}
But this approach essentially uses the equation of state of the microscopic
model and does not make use of the CG model with DD potentials. 

It is also possible to derive expressions for the CG 
pressure and chemical potential
by using the active approach of Ref.~\onlinecite{SST-02}.
The idea is to start from Eq.~(\ref{FCG-def}), set $\rho_p = \rho = N/V$,
and then take the derivative with respect to $V$.
This gives the well-known 
Ascarelli-Harrison\cite{AH-69} expression for the pressure:
\begin{eqnarray}
P_{AH}(\rho) &=& \rho^2 {\partial f_{CG}(\rho,\rho)\over \partial \rho} 
\nonumber \\
   &=& P_{CG}(\rho,\rho) + \rho^2 \left(
    {\partial f_{CG}(\rho,\rho_p)\over \partial \rho_p}\right)_{\rho_p = \rho} 
\nonumber \\
&=& P_{\rm vir}(\rho,\rho) + 
 \\
   && \quad 2 \pi \rho^3 
   \int_0^\infty r^2 dr\, {\partial V_{D,CG}({\bf r};\rho)\over \partial \rho} 
    g_{CG}({\bf r};\rho,\rho),
\nonumber 
\end{eqnarray}
where $P_{\rm vir}(\rho,\rho)$ is explicitly given in
Eq.~(\ref{Virialpressure}).
Simple algebra allows us to rewrite 
\begin{equation}
\beta P_{AH}(\rho) = 
    \int_0^\rho d\sigma\, 
    \left[K_{CG}(\sigma,\rho) + {\rho(\rho-\sigma)\over \sigma} 
     {\partial K_{CG}(\sigma,\rho) \over \partial \rho}\right],
\end{equation}
which shows that $P_{AH}(\rho)$ differs in general from 
the pressure of the underlying model and that its density derivative differs
from the CG quantity $K_{CG}(\rho,\rho)$. 

The Ascarelli-Harrison prescription can also be applied to the chemical 
potential, defining
\begin{eqnarray}
\mu_{AH}(\rho) &=& f_{CG}(\rho,\rho) + 
     \rho {\partial f_{CG}(\rho,\rho)\over \partial \rho} = \nonumber \\
    &=& \mu_{CG}(\rho,\rho) + \rho 
      \left( {\partial f_{CG}(\rho,\rho_p) \over \partial \rho_p}
      \right)_{\rho_p = \rho} = \nonumber \\
     &=&
   \mu_{CG}(\rho,\rho) + \label{muAH} \\
    && \quad 2 \pi \rho^2 
   \int_0^\infty r^2 dr\, {\partial V_{D,CG}({\bf r};\rho)\over \partial \rho} 
    g_{CG}({\bf r};\rho,\rho).
\nonumber
\end{eqnarray}
Again, it is easy to verify that such an expression differs from
the underlying-model result $\beta \mu^{\rm exc}(\rho)$. Indeed, we have 
\begin{equation}
\beta \mu_{AH}^{\rm exc}(\rho) = 
    \int_0^\rho {d\sigma\over \sigma}\, 
    \left[K_{CG}(\sigma,\rho) - 1 + {(\rho-\sigma)} 
     {\partial K_{CG}(\sigma,\rho) \over \partial \rho}\right].
\end{equation}
As we have seen, the passive and the active approaches provide two different
expressions for the pressure. It is interesting to reinterpret these results 
as follows.
For standard thermodynamic systems there are two equivalent approaches to the 
pressure. One can use the force route, in which one defines the pressure 
as the force per unit area exerted on the walls of the box containing the 
system. For a system of $N$ molecules in a volume $V$ 
interacting with a pair potential $v({\bf r})$, Clausius theorem allows one 
to relate the pressure $P$ so defined to the virial:
\cite{HansenMcDonald}
\begin{equation}
PV = N k_B T - {1\over 3} \left\langle 
   \sum_{i<j} {\bf r}_{ij} \cdot {\partial v({\bf r}_{ij})\over {\bf r}_{ij}}
\right\rangle .
\label{Clausius}
\end{equation}
Alternatively, one can define the pressure $P$ by using the work $dW$ 
necessary to change the volume of the box by an infinitesimal amount $dV$,
$dW = -P dV$, which provides the usual expressions of $P$ as 
the derivative of the thermodynamic potentials with respect to the volume.
The two definitions are obviously equivalent for density-independent potentials 
but differ in the presence of DD interactions. 
In the passive approach the pressure satisfies Clausius theorem. 
This choice looks very natural, since 
the Clausius equation (\ref{Clausius})
is a simple mechanical relation that is proved by balancing the forces 
for a system of a {\em fixed} number of particles in a box of 
{\em fixed} volume, hence at fixed density. The density dependence of the
potential is irrelevant and therefore,
one would naturally expect Eq.~(\ref{Clausius})
to hold also in the presence of DD interactions.
On the other hand, $P_{AH}(\rho)$ is defined by using the 
thermodynamic route, i.e. it satisfies the usual thermodynamic 
relation $dW = -P_{AH} dV$. However, it is not consistent with Clausius
theorem, hence the pressure computed in the active approach is not directly
related to the force exterted by the particles. This is due to the fact 
that the work done when changing the volume of the system has two
contributions. One of them corresponds to the work done against the forces 
acting 
on the box boundary (this is the only contribution taken into account
when considering the virial pressure), 
the other one takes into account the change of the potential as the volume 
$V$ is changed at fixed number of particles. From this viewpoint, the 
active approach certainly looks less natural than the passive one. 

\section{Comparing the different definitions of the pressure and chemical
potential in the canonical ensemble} \label{sec3}

As we have seen in the previous section, the DD quantities
$P_{CG}(\rho,\rho)$ and $\mu_{CG}(\rho,\rho)$ (passive approach) 
or their counterparts $P_{AH}(\rho)$ and $\mu_{AH}(\rho)$ only 
provide approximations
to the correct result. In this section we wish to compare them 
with the exact result, determining quantitatively the size of the 
discrepancy. In the context of soft-matter systems,
\cite{Likos-01,DPP-12-a,DPP-12-b,Pierleoni:2007p193,Vettorel:2010p1733}
it is also common
to use CG models based on zero-density potentials, i.e. potentials obtained by
matching the pair distribution function in the limit $\rho \to 0$. 
In this approach the potentials are fixed and then used to obtain predictions
at densities $\rho > 0$. Hence, in the notation of the previous 
section, the pressure and the chemical potential correspond to 
$P_{CG}(\rho,0)$ and $\mu_{CG}(\rho,0)$, respectively.
These quantities  will be compared
with the corresponding DD quantities, in order to understand which 
method provides the best approximation.
We shall first consider the low-density limit, which can be analyzed 
in a model-independent way, and then 
we shall apply all formulae to a specific soft-matter example, polymers in the 
semidilute regime. In both cases, the results of the different approaches 
will be compared with those obtained in the microscopic model.

\subsection{Low-density behavior} \label{sec3.1}

Let us consider the low-density limit. For $\rho\to 0$ the $XX$ 
pair distribution function can be expanded as 
\begin{eqnarray}
g({\bf r};\rho) = g_0({\bf r}) + \rho g_1({\bf r}) + O(\rho^2).
\label{grho-lowdens}
\end{eqnarray}
We define $h_{0}({\bf r}) = g_0({\bf r}) - 1$,
\begin{equation}
\hat{g}_1({\bf r}) = g_0({\bf r}) \int d^3{\bf s}\, h_0({\bf s}) h_0({\bf r-s}),
\end{equation}
and $\Delta({\bf r}) = g_1({\bf r}) - \hat{g}_1({\bf r})$. The quantity 
$\Delta({\bf r})$ encodes the contributions of the three-body interactions:
for monoatomic systems interacting by means of pairwise additive interactions 
we have 
$\Delta({\bf r}) = 0$
and $g_1({\bf r}) = \hat{g}_1({\bf r})$.
Using the compressibility relation (\ref{compr_rel}) and 
Eq.~(\ref{grho-lowdens}), we can compute $K(\rho)$ in the underlying 
microscopic model:
\begin{equation}
K(\rho) = 1 - \rho I_0 - \rho^2 (I_1 + I_2) + O(\rho^3),
\label{K-expand-FA}
\end{equation}
with 
\begin{eqnarray}
I_0 &=& \int d^3{\bf r}\, h_0({\bf r}), \\
I_1 &=& \int d^3{\bf r} d^3{\bf s}\, h_0({\bf r})h_0({\bf s}) h_0({\bf r-s}),
\\
I_2 &=& \int d^3{\bf r}\, \Delta({\bf r}).
\end{eqnarray}
Eqs.~(\ref{pFA-KFA}) and (\ref{muFA-KFA}) give
\begin{eqnarray}
\beta P(\rho) &=& \rho - {1\over2} \rho^2 I_0 - {1\over3} \rho^3 (I_1 + I_2) + 
     O(\rho^4) , \\
\beta \mu^{\rm exc}(\rho) &=& - I_0 \rho - {1\over2} \rho^2 (I_1 + I_2) + 
     O(\rho^3).
\end{eqnarray}
Let us now consider the CG model. First, we need to compute the 
DD potential $V_{D,CG}({\bf r};\rho_p)$. 
Since we are interested in the low-density limit, 
we expand it as 
\begin{equation}
  V_{D,CG}({\bf r};\rho_p) = V_{0CG}({\bf r}) + \rho_p V_{1CG}({\bf r}) + 
  O(\rho_p^2).
\end{equation}
At leading order in the density, we have 
\begin{equation}
  g_{CG}({\bf r};\rho,\rho_p) = e^{-\beta V_{0CG}({\bf r})} + 
    O(\rho,\rho_p).
\end{equation}
By requiring the equality of the pair distribution functions, i.e. 
$g({\bf r};\rho) = g_{CG}({\bf r};\rho,\rho)$, we obtain 
\begin{eqnarray}
\beta V_{0CG}({\bf r}) &=& - \ln g_0({\bf r}).
\label{V0CG}
\end{eqnarray}
If we include the first density correction we have 
\begin{eqnarray}
&& g_{CG}({\bf r};\rho,\rho_p) = 
    e^{-\beta V_{D,CG}({\bf r};\rho_p)}
    \left[1 + 
    \vphantom{\left(e^{-\beta V_{D,CG}({\bf s};\rho_p)} - 1\right)}
   \right.
\\ 
&& \qquad \left.
    \rho \int d^3{\bf s} f({\bf s};\rho_p) 
    f({\bf s} - {\bf r};\rho_p)\right] + O(\rho^2),
\nonumber 
\end{eqnarray}
where 
\begin{equation}
f({\bf s};\rho_p) = e^{-\beta V_{D,CG}({\bf s};\rho_p)} - 1.
\nonumber
\end{equation}
Expanding in $\rho_p$, using Eq.~(\ref{V0CG}) and 
$h_0({\bf r}) = g_0({\bf r}) - 1$, we obtain
\begin{eqnarray}
&& g_{CG}({\bf r};\rho,\rho_p) =
   g_0({\bf r}) \left[1 - \rho_p \beta V_{1CG}({\bf r}) \right] \nonumber \\
&& \qquad + 
    \rho g_0({\bf r})
     \int d^3{\bf s} \,
   h_0({\bf s}) h_0({\bf r}-{\bf s}) + O(\rho^2,\rho\rho_p,\rho_p^2).
\end{eqnarray}
By requiring 
$g({\bf r};\rho) = g_{CG}({\bf r};\rho,\rho)$, we obtain 
the next correction:
\begin{eqnarray}
\beta V_{1CG}({\bf r}) &=& - {\Delta({\bf r})\over g_0({\bf r})}.
\end{eqnarray}
Note that these definitions imply
\begin{equation}
\int d^3{\bf r}\, f({\bf r};\rho_p) = 
  I_0 + \rho_p I_2 + O(\rho_p^2),
\label{eq:25}
\end{equation}
for $\rho_p \to 0$.
We can now  compute $K(\rho,\rho_p)$. 
In the limit $\rho\to 0$, using the standard expressions valid 
for a monoatomic system,\cite{HansenMcDonald} we obtain
\begin{eqnarray}
&& K_{CG}(\rho,\rho_p) = 1 - 
   \rho \int d^3{\bf r}\, f({\bf r};\rho_p)
\\
&& \quad -
   \rho^2 \int d^3{\bf r} d^3{\bf s}\, 
  f({\bf r};\rho_p) f({\bf s};\rho_p)
  f({\bf r}-{\bf s};\rho_p) + O(\rho^3).
\nonumber 
\end{eqnarray}
Expanding in $\rho_p$ and using Eq.~(\ref{eq:25}), we obtain
\begin{equation}
K_{CG}(\rho,\rho_p) = 
  1 - \rho I_0 - \rho^2 I_1 - \rho \rho_p I_2 + 
  O(\rho^3, \rho^2 \rho_p, \rho \rho_p^2).
\end{equation}
For $\rho_p = \rho$ this expression coincides with Eq.~(\ref{K-expand-FA}),
as expected. Using Eqs.~(\ref{muCG-K}), (\ref{pCG-K}), and (\ref{fCG-K}) 
we obtain
\begin{eqnarray}
\beta P_{CG}(\rho,\rho_p) &=&
   \rho - {1\over2} \rho^2 I_0 - {1\over 3} \rho^3 I_1 - 
    {1\over 2} \rho^2 \rho_p I_2,  \\
\beta \mu^{\rm exc}_{CG}(\rho,\rho_p) &=&
   -\rho I_0 - {1\over 2} \rho^2 I_1 - \rho \rho_p I_2, \\
\beta f_{CG}(\rho,\rho_p) &=&
  \ln \rho - 1 - {1\over2} \rho I_0 - {1\over 6} \rho^2 I_1 - 
   {1\over2} \rho \rho_p I_2,
\end{eqnarray}
disregarding higher-order terms in $\rho$ and $\rho_p$.
From the expression of the free energy, we obtain
the Ascarelli-Harrison expressions for the pressure and 
chemical potential:
\begin{eqnarray}
\beta P_{AH}(\rho) &=& \beta \rho^2 {df_{CG}(\rho,\rho)\over d\rho} =
   \rho - {1\over2} \rho^2 I_0 - {1\over3} \rho^3 I_1 - \rho^3 I_2, \\
\beta \mu^{\rm exc}_{AH}(\rho) &=&
   \beta f_{CG}(\rho,\rho) + \beta \rho {df_{CG}(\rho,\rho)\over d\rho} 
\nonumber \\
   &=&
   -\rho I_0 - {1\over 2} \rho^2 I_1 - {3\over2} \rho^2  I_2. 
\end{eqnarray}
We can thus compare these expressions with the exact ones valid for the 
underlying model. For the pressure 
we have
\begin{eqnarray}
&& \beta P_{CG}(\rho,\rho) - \beta P(\rho) = 
    -{1\over 6} \rho^2 I_2, \nonumber \\
&& \beta P_{CG}(\rho,0) -\beta P(\rho) = 
     {1\over 3} \rho^2 I_2, \nonumber \\
&& \beta P_{AH}(\rho) - \beta P(\rho) = 
    -{2\over3} \rho^2 I_2.
\label{deltap-lowdens-can}
\end{eqnarray}
From these results we see that the pressure computed by using the 
DD potentials provides the best approximation to $P(\rho)$. The 
AH formula is significantly worse---deviations are four times larger than those
for the virial pressure $P_{CG}(\rho,\rho)$---as already noted 
in Refs.~\onlinecite{Louis-02,JHGL-07}. By 
using the zero-density potentials one obtains a result which is worse by a
factor of two than the DD result.

We can perform the same comparison for the chemical potential. We find
\begin{eqnarray}
&& \beta \mu_{CG}^{\rm exc}(\rho,\rho) - \beta \mu^{\rm exc}(\rho) = 
   -{1\over2} \rho^2 I_2, \\
&& \beta \mu_{CG}^{\rm exc}(\rho,0) - \beta \mu^{\rm exc}(\rho) = 
   {1\over2} \rho^2 I_2, \\
&& \beta \mu_{AH}^{\rm exc}(\rho) - \beta \mu^{\rm exc}(\rho) = 
   -\rho^2 I_2. 
\end{eqnarray}
In this case, the chemical potentials derived by using the 
DD and the density-independent potentials
have the same accuracy. 
Expression (\ref{muAH}) is instead worse than both the one obtained by using
DD potentials and the one obtained by using zero-density potentials.

Note that all deviations are proportional to $I_2$, which encodes the 
contributions of the (effective) three-body interactions at this order in
$\rho$. Hence, 
consistency is only possible if the three-body terms 
do not contribute to the thermodynamics, hence $I_2 = 0$,
a result which is not generally true since the free energy of the 
CG model in a generic CG procedure is never pairwise additive.

\subsection{Polymers in the semidilute regime} \label{sec3.2}

\begin{table*}
\caption{Compressibility factor $Z = \beta P/\rho$ 
and excess chemical potential for a polymer 
CG system at $\Phi = 0.8$ and $1.5$. The reference polymer $Z(\rho)$ and 
$\mu^{\rm exc}(\rho)$ have been obtained 
by using the equation of state reported 
in Ref.~\protect\onlinecite{Pelissetto-08}. 
}
\label{tableDD}
\begin{tabular}{cccccccc}
\hline\hline
$\Phi$ & $Z(\rho)$ & $Z_{CG}(\rho,0)$ & $Z_{CG}(\rho,\rho)$ &
     $Z_{AH}(\rho)$ & $\beta \mu^{\rm exc}(\rho)$ &
     $\beta \mu^{\rm exc}_{CG}(\rho,0)$ & 
     $\beta \mu^{\rm exc}_{CG}(\rho,\rho) $ \\
\hline
0.80 & 2.35 & 2.22402(4) & 2.44790(4) & 2.51(2)& 2.56 & 2.3797(2) & 2.8108(3) \\
1.50 & 3.90 & 3.39847(4) & 4.13973(4) & 4.28(5)& 5.38 & 4.653(4)  & 6.0929(4) \\
\hline\hline
\end{tabular}
\end{table*}

To assess the quality of the different approximations to the pressure and the 
chemical potential, we study the thermodynamic behavior of 
a system of linear polymers under good-solvent conditions. 
We have extensively studied CG models for this type of systems in
Refs.~\onlinecite{DPP-12-a,DPP-12-b}. 
For two values of the polymer volume fraction $\Phi$,
\begin{eqnarray}
  \Phi = {4\pi\over 3} \hat{R}_g^3 \rho,
\end{eqnarray}
where $\hat{R}_g$ is the zero-density radius of gyration, namely
$\Phi = 0.80$ and $\Phi = 1.50$, we compute \cite{footnote1} the 
center-of-mass pair distribution function $g({\bf r};\rho)$. 
Then, by using the method described in 
Refs.~\onlinecite{LBHM-00,BL-02}, 
we determine the DD potential $V_{D,CG}({\bf r};\rho)$,
checking that it reproduces the full-monomer value of $K$. 
This is a strong check of the accuracy with which the potential is
determined, since the compressibility is very sensitive to the large-distance
behavior of the pair potentials.\cite{MullerPlathe-02}
For the zero-density potential we use the accurate 
expression reported in Ref.~\onlinecite{PH-05}. The Monte Carlo 
estimates of the 
compressibility factor $Z = \beta P/\rho$ 
and of the chemical potential obtained by using the 
virial route and Widom's method, respectively, are reported in 
Table~\ref{tableDD}. Completely consistent results are obtained by using the 
integral-equation approach and the hypernetted-chain (HNC) closure.
In our previous work \cite{DPP-12-a} 
we showed that the compressibility factor 
$Z_{CG}(\rho,0)$ obtained by using zero-density potentials
underestimates the polymer $Z(\rho)$, which, in the low-density limit,
implies $I_2 < 0$. Our present results are fully consistent.
Differences increase with 
increasing $\Phi$, as expected: the relative error is 5\% and 13\% for 
$\Phi = 0.8$ and 1.5, respectively. If one uses the DD potentials 
the compressibility factor is overestimated, by 4\% and 6\% in the two cases.
Therefore, in agreement with the low-density analysis, DD potentials 
provide more accurate estimates of the pressure and clearly represent 
an improvement with respect to zero-density potentials, especially for 
the largest value of $\Phi$. We also compute the pressure by 
using the Ascarelli-Harrison expression. A precise determination is not easy:
indeed, it is difficult to estimate accurately
the derivative $\partial V_{D,CG}({\bf r};\rho)/\partial \rho$, especially
for $r/R_g\gtrsim 2$. Hence, the additional term that appears in the 
Ascarelli-Harrison expression can only be determined with limited
precision, which we estimated somewhat subjectively by looking
at the variation of the results when different parametrizations of the 
DD potentials are used: we estimate a relative error on the correction term 
of  approximately 20-30\%. As expected, $Z_{AH}$ is significantly
worse than the virial expression $Z_{CG}(\rho,\rho)$: the relative error is 
7\% and 12\% for the two values of $\Phi$. We also considered the combination
$P'(\rho) = {4\over3} P_{CG}(\rho,\rho) - {1\over3} P_{AH}(\rho)$,
which is correct to order $\rho^2$, see Eq.~(\ref{deltap-lowdens-can}). 
We find $Z'(\rho) = 2.43(1)$, 4.06(2) for the two values of $\Phi$.
These estimates are slightly closer to the exact result than 
$Z_{CG}(\rho,\rho)$. 

Similar conclusions are reached for the chemical potential. If one uses 
zero-density potentials, $\beta \mu^{\rm exc}(\rho)$ is underestimated
by 7\% and 14\% in the two cases, while, by using DD potentials
one overestimates the chemical potential by 9\% and 13\%. As already discussed
in the low-density regime, zero-density and DD potentials are equally
inaccurate. 

This discussion indicates that, for $\Phi\lesssim 1$, which is the region 
in which many-body effects are small \cite{Pierleoni:2007p193} and therefore
CG monoatomic models should work reasonably well, there is little
advantage in using DD potentials. Accurate results can only be obtained 
by using multiblob models.\cite{Pierleoni:2007p193,Fritz:2009p1721,%
CG-10,Vettorel:2010p1733,DPP-12-a,DPP-12-b} 
For instance, the tetramer model (model 4MB-2) with zero-density potentials
of Ref.~\onlinecite{DPP-12-b} gives $Z(\rho) = 2.3505(4)$ and 
$\beta \mu^{\rm exc}(\rho)=2.57(1)$ for $\Phi = 0.8$, in perfect agreement 
with the full-monomer results. For $\Phi = 1.5$, small deviations,
of the order of 2-3\%, are found: $Z = 3.7901(2)$ and
$\beta \mu^{\rm exc} = 5.27(2)$. If more accuracy is needed, one can 
increase the number of blobs. If one uses a decamer with 10
blobs,\cite{DPP-12-b}
one obtains for $\Phi = 1.5$ the estimates $Z = 3.929(3)$
and $\beta \mu^{\rm exc} = 5.3(1)$, which are consistent with the 
full monomer results.

\section{Fugacity-dependent potentials in the grand-canonical ensemble} 
\label{sec4}

The previous discussion applies to the canonical ensemble. However,
if one is interested  in studying phase coexistence, the grand canonical
ensemble is the natural one. In order to implement the CG 
procedure, we parametrize all thermodynamic variables in terms of the 
reduced fugacity $z = q_1 e^{\beta\mu}$, where $q_1 = Z_1/V$ and 
$Z_1$ is the partition function of a single molecule. This guarantees
that $z\approx \rho$ for $\rho\to 0$, both in the original and in the 
CG model. As before, the basic quantity in the approach is the 
$XX$ pair correlation function $g({\bf r};z)$, which is now a function
of the fugacity $z$. If $\rho(z)$ gives the density of the 
underlying system as a function of $z$,
we have obviously $g({\bf r};z) = g({\bf r};\rho(z))$. The relation with the 
thermodynamics is provided by the compressibility relation.  In
the grand canonical ensemble the role of $K$ is played by
\begin{equation}
   H(z) = - z {\partial \over \partial z} \left( {z\over \rho}\right) = 
   {z\over \rho} \left( {\langle N^2\rangle - \langle N\rangle^2 \over 
         \langle N\rangle} - 1\right) ,
\end{equation}
which satisfies\cite{HansenMcDonald}
\begin{equation}
   H(z) = z \int d^3{\bf r}\, [g({\bf r};z) - 1].
\label{compr-rul-H}
\end{equation}
Using the standard thermodynamic relations we obtain 
(of course, we assume here that there is no phase transition in the 
interval $[0,z]$)
\begin{eqnarray}
&&\rho(z) = z \left[1  - \int_0^z {dw\over w}\, H(w)\right]^{-1} ,
\label{rho-GC-FA} \\
&&\beta P(z) = \int_0^z {dw\over w}\, \rho(w) .
\label{p-GC-FA}
\end{eqnarray}
In the CG model one considers a fugacity-dependent (FD) pair potential 
$V_{F,CG}({\bf r};z)$. Again, we should distinguish between the passive and
the active approach. For this purpose, we start by considering
the fugacity appearing in the 
potential as a parameter and define 
\begin{equation}
\beta \Omega_{CG}(z,z_p) = - \ln \sum_{N=0}^\infty {z^N\over N!} Q_N(z_p),
\label{OmegaCG}
\end{equation}
where $Q_N(z_p)$ is the partition function of a system of $N$ monoatomic
molecules interacting via $V_{F,CG}({\bf r};z_p)$. 
The FD potential $V_{F,CG}({\bf r};z_p)$ is fixed so that the pair
distribution function in the CG model $g_{CG}({\bf r};z,z_p)$ for $z_p = z$ 
is equal to the $XX$ pair distribution function in the underlying model. 
At fixed $z_p$, we can define the standard thermodynamic quantities:
\begin{eqnarray}
 P_{CG}(z,z_p) &=& - {1\over V} \Omega_{CG}(z,z_p), \\
 \rho_{CG}(z,z_p) &=&  -{\beta z\over V} 
    \left({\partial\Omega_{CG}\over \partial z} \right)_{T,V,z_p}.
\end{eqnarray}
Since $z_p$ is a fixed parameter,
the density $\rho_{CG}(z,z_p)$ can be computed as $\langle N\rangle_{z,z_p}/V$
as usual [here $\langle \cdot\rangle_{z,z_p}$ is the average with respect
to the grand canonical distribution defined by Eq.~(\ref{OmegaCG})], 
while $P_{CG}(z,z_p)$ can be determined by using the 
grand-canonical generalization of the virial expression. If the
pair distribution function $g_{CG}({\bf r};z,z_p)$ is known, we can 
also compute thermodynamic quantities using the compressibility route. 
First, we determine
\begin{equation}
   H_{CG}(z,z_p) = z \int d^3{\bf r}[g_{CG}({\bf r};z,z_p) - 1],
\end{equation}
then $\rho_{CG}(z,z_p)$
and $\beta P_{CG}(z,z_p)$ are derived by using Eqs.~(\ref{rho-GC-FA}) 
and (\ref{p-GC-FA}) at fixed $z_p$. 
The equality of the pair distribution functions implies
$H_{CG}(z,z) = H(z)$, but this result does not extend to $\rho_{CG}(z,z)$ and 
$P_{CG}(z,z)$. For instance,
\begin{eqnarray}
&& {z\over \rho_{CG}(z,z)} = 1 - \int_0^z {dw\over w} H_{CG}(w,z) 
\nonumber \\ && \qquad \not= 
{z\over \rho(z)} = 1 - \int_0^z {dw\over w} H_{CG}(w,w).
\label{zoverrho-comparison}
\end{eqnarray}
Let us now consider the CG model at fugacity $z=z_p$. To define the pressure 
there is no need to distinguish between the active and the passive approach,
since $P$ is directly related to $\Omega$. Hence, whatever approach 
is used, the pressure is always given by $P_{CG}(z,z)$ and therefore 
always satisfies Clausius theorem. For the density 
instead, the two approaches give different results. In the passive 
approach, we should consider $\rho_{CG}(z,z) = \langle N\rangle_{z,z}/V$,
while in the active approach we should consider 
\begin{eqnarray}
\rho_{\rm active}(z) &=& - {\beta z \over V} 
      {\partial \Omega_{CG}(z,z) \over \partial z} 
\nonumber \\
     &=& {\langle N\rangle_{z,z}\over V} - 
    \left.    {\beta z \over V} 
    {\partial \Omega_{CG}(z,z_p) \over \partial z_p}
    \right|_{z=z_p}.
\end{eqnarray}
This definition looks quite unnatural, since
$\rho_{\rm active}(z)$ differs from the average 
number of particles per unit volume present in the system.
    
It is important to stress that the DD potential $V_{D,CG}({\bf r};\rho)$ 
defined in the canonical ensemble and the FD potential
$V_{F,CG}({\bf r};z)$ are not simply related. If $\rho(z)$ gives the 
density as a function of $z$ in the underlying model, $V_{F,CG}({\bf r};z)$ 
differs from $V_{D,CG}({\bf r};\rho(z))$: the same thermodynamic
state in the microscopic model is represented by 
{\em different}\/ state-dependent
potentials, that depend on the ensemble one chooses. Ensemble equivalence
does not hold for state-dependent potentials. 
To understand why the equality does not hold, note that the system at fugacity 
$z$ interacting with pair potential $V_{F,CG}({\bf r};z)$ has
density $\rho_{CG}(z,z)\not=\rho(z)$,
see Eq.~(\ref{zoverrho-comparison}).
Hence, $V_{F,CG}({\bf r};z)$ and $V_{D,CG}({\bf r};\rho(z))$ correspond to 
two systems that have the same pair distribution function but at {\em different}
densities: hence they cannot be equal. The only exception is the 
zero-density/zero-fugacity case. Since $\rho_{CG}(0,0) = 0$, we have 
$V_{F,CG}({\bf r};z=0) = V_{D,CG}({\bf r};\rho=0)$.
The ensemble inequivalence, which is 
completely general and extends to other ensembles, 
say the isothermal-isobaric one,
is clearly related to the thermodynamic inconsistency of state-dependent
potentials, which are not able to reproduce simultaneously the correct 
value of density, chemical potential, and pressure. It 
is also completely consistent with Henderson's theorem,\cite{Henderson-74}
which states that, if two systems interacting by pairwise potentials
have the same pair distribution function at the {\em same} density, 
then the pair potential is unique. Indeed, in the different ensembles,
the pair distribution function is the same, but the density is not. 

\section{Comparing the estimates of the density and pressure in the 
grand canonical ensemble} \label{sec5}

We shall now repeat the analysis of Sec.~\ref{sec3} in the grand-canonical 
ensemble, comparing the exact $\rho(z)$ and $P(z)$ with the 
zero-density approximations $\rho_{CG}(z,0)$ and $P_{CG}(z,0)$ and 
the approximations $\rho_{CG}(z,z)$ and $P_{CG}(z,z)$ obtained by using 
FD potentials. We will first consider the low-fugacity 
limit and then we will give results for a specific example, polymers under 
good-solvent conditions in the semidilute regime.

\subsection{Low-fugacity limit}

Since $g({\bf r};z) = g({\bf r};\rho(z))$ and we have $z = \rho + O(\rho^2)$
for $\rho\to 0$, we can reexpress Eq.~(\ref{grho-lowdens}) as 
\begin{equation}
g({\bf r};z) = g_0({\bf r}) + z [\hat{g}_1({\bf r}) + \Delta({\bf r})] + O(z^2).
\end{equation}
Using Eq.~(\ref{compr-rul-H}) we obtain
\begin{equation}
H(z) = z I_0 + z^2 (I_0^2 + I_1 + I_2) + O(z^3),
\end{equation}
where $I_0$, $I_1$, and $I_2$ are defined in Sec.~\ref{sec2}.
Eqs.~(\ref{rho-GC-FA}) and (\ref{p-GC-FA}) give 
\begin{eqnarray}
\rho(z) &=& z + I_0 z^2 + {1\over2} z^3 (3 I_0^2 + I_1 + I_2) + O(z^4) \\
\beta P(z) &=& z + {1\over2} I_0 z^2 + 
     {1\over 6} z^3 (3 I_0^2 + I_1 + I_2) + O(z^4). 
\end{eqnarray}
Analogously, in the CG model we obtain
\begin{eqnarray}
H_{CG}(z,z_p) = z I_0 + z^2 (I_0^2 + I_1) + z z_p I_2 + O(z^3, z^2 z_p, z
z_p^2), 
\end{eqnarray}
which satisfies $H_{CG}(z,z) = H(z)$. From this expression we obtain
\begin{eqnarray}
\rho_{CG}(z,z_p) &=& z + z^2 I_0 + {1\over 2} z^3 (3 I_0^2 + I_1) + 
   z^2 z_p I_2, \\
\beta P_{CG}(z,z_p) &=& z + {1\over2} z^2 I_0 + 
     {1\over 6} z^3 (3 I_0^2 + I_1) + 
     {1\over 2} z^2 z_p I_2,
\end{eqnarray}
disregarding higher-order terms in $z$ and $z_p$. For the density in the 
active approach we obtain instead
\begin{equation}
\rho_{\rm active}(z) = z + z^2 I_0 + {1\over 2} z^3 (3 I_0^2 + I_1 + 3 I_2)  .
\end{equation}
Comparing the CG results with the exact ones we obtain for the density
\begin{eqnarray}
\rho_{CG}(z,z) - \rho(z) &=& {1\over2} I_2 z^3, \\
\rho_{CG}(z,0) - \rho(z) &=& -{1\over2} I_2 z^3, \\
\rho_{\rm active}(z) - \rho(z) &=& I_2 z^3.
\end{eqnarray}
In this case $\rho_{CG}(z,0)$ and $\rho_{CG}(z,z)$ provide 
approximations that differ by the same amount from the correct,
underlying-model result. As in the case of DD potential, the density 
computed in the active approach is instead significantly worse.
If we consider the pressure we obtain
\begin{eqnarray}
\beta P_{CG}(z,z) - \beta P(z) &=& {1\over3} I_2 z^3, \\
\beta P_{CG}(z,0) - \beta P(z) &=& -{1\over6} I_2 z^3.
\end{eqnarray}
It is somewhat surprising that in this case zero-fugacity potentials
are more accurate than those that depend on the fugacity.  We can also
compute the compressibility factor $Z = \beta P/\rho$, obtaining
\begin{eqnarray}
Z_{CG}(z,z) - Z(z) &=& -{1\over 6} I_2 z^2, \\
Z_{CG}(z,0) - Z(z) &=& {1\over 3} I_2 z^2. 
\end{eqnarray}
As in the case of the canonical ensemble, the compressibility factor 
is instead more accurate for state-dependent potentials.

\subsection{Polymers in the semidilute regime} \label{sec5.2}

We will now repeat in the grand canonical ensemble the computations
we have reported in Sec.~\ref{sec3.2}. 
We consider the two values of
$z$, $z\hat{R}_g^3 = 2.47$ and 77.7 that correspond to 
$\Phi = 0.8$ and $\Phi = 1.5$, the values of the polymer 
volume fraction considered in Sec.~\ref{sec3.2}. To compute the 
FD potentials---as we already noted they differ from the 
DD potentials---we use the same technique employed in 
Ref.~\onlinecite{LBHM-00,BL-02}. Since we expect the HNC
approximation to be quite accurate\cite{HansenMcDonald}
(we verify it below),
we relate $z$ and $\rho$ by using 
the HNC expression \cite{Morita-60,Attard-91}
\begin{equation}
z = \rho\exp\left[ {\rho\over2} 
   \int d^3{\bf r} \left( h({\bf r})^2 - h({\bf r}) c({\bf r}) - 
    2 c({\bf r}) \right)\right],
\label{zHNC}
\end{equation}
where $h({\bf r}) = g({\bf r};z) - 1$ and the direct correlation function 
$c({\bf r})$ is given by the Ornstein-Zernike relation \cite{HansenMcDonald}
\begin{equation}
h({\bf r}) = c({\bf r}) + \rho \int d^3{\bf s}\, 
  c({\bf s}) h({\bf r} - {\bf s}).
\label{OZ}
\end{equation}
Given $z$ and $g({\bf r};z)$, we can solve Eqs.~(\ref{zHNC}) and 
(\ref{OZ}) to obtain $\rho$ and $c({\bf r})$. The potential follows 
from the HNC closure relation:
\begin{equation}
\beta V_{F,CG}({\bf r};z) = h({\bf r}) - c({\bf r}) - 
     \ln g({\bf r};z).
\label{VFCG-HNC}
\end{equation}
In this formalism it is easy to understand why the FD
potential at, say, $z\hat{R}_g^3 = 2.47$ is different from the DD 
potential at $\Phi = 0.8$, although $z\hat{R}_g^3 = 2.47$ and $\Phi = 0.8$
correspond to the same thermodynamic state of the underlying model. Indeed,
Eqs.~(\ref{zHNC}) and (\ref{OZ}) cannot be simultaneously satisfied 
by fixing the chemical potential and the density to the full-monomer 
values. In the DD case, one fixes $\rho$, obtaining a
fugacity that differs from the full-monomer value, while here we fix $z$, 
obtaining a different density. 
The HNC procedure can also be applied to other ensembles. For instance,
one could obtain the potential in the isobaric ensemble, by simultaneously
solving Eqs.~(\ref{OZ}), (\ref{VFCG-HNC}), and (\ref{Virialpressure}),
fixing the pressure to its full monomer value. Of course, both $\rho$ and 
$z$ would differ from the corresponding full-monomer values. 
\begin{figure*}[t]
\begin{center}
\begin{tabular}{c}
\epsfig{file=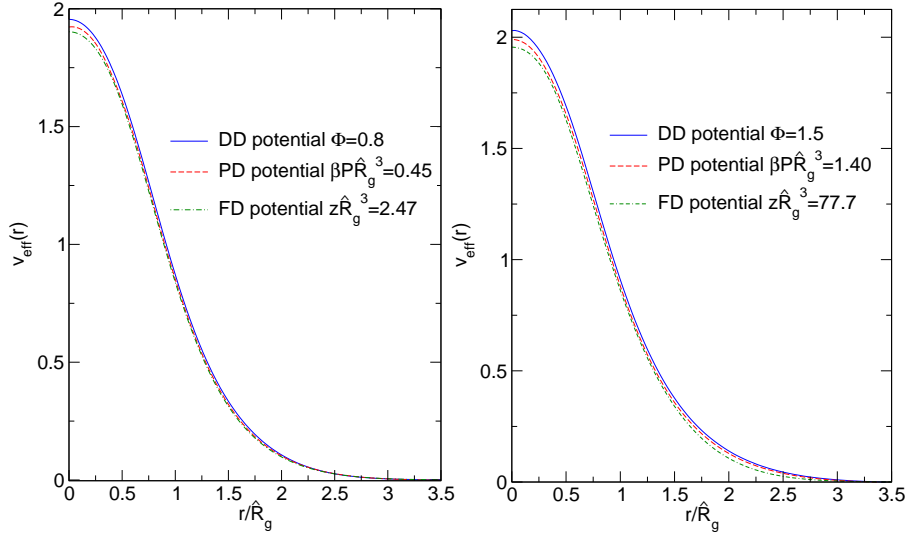,angle=0,width=12truecm} \hspace{0.5truecm} \\
\end{tabular}
\end{center}
\caption{We report the DD potential (DD), the fugacity dependent (FD) 
potential, and the pressure dependent (PD) potential as a function 
of $r/\hat{R}_g$. They are
obtained by fixing: 
(left) $\Phi = 0.80$, $z\hat{R}_g^3 = 2.47$, 
$\beta P\hat{R}_g^3 = 0.45$ in the three cases, respectively;
(right) $\Phi = 1.50$, $z\hat{R}_g^3 = 77.7$, 
$\beta P\hat{R}_g^3 = 1.40$, respectively.
}
\label{fig:potentials}
\end{figure*}
The state dependent potentials are reported in Fig.~\ref{fig:potentials}
for the two different thermodynamic points of the polymer system we have 
considered. We also report the potential in the isobaric ensemble,
obtained by fixing the pressure.\cite{fixpress}
Differences among the potentials are not
large on the scale of the figure, but still not negligible. 
At full overlap $r=0$, we have\cite{PH-05} 
$\beta V_{CG}(r=0) = 1.775(5)$ for the 
zero-density potential (note that this potential is independent of the
ensemble). At the state point with $\Phi=0.8$ we have 
$\beta V_{F,CG}(r=0) \approx 1.902$ for the FD potential and 
$\beta V_{D,CG}(r=0) \approx 1.955$ for the DD potential. Hence, 
in the grand canonical
ensemble many-body interactions increase
repulsion by 8\% with respect to the zero-density case. In the 
canonical ensemble the effect is 40\% larger: repulsion increases by 11\%.
At the larger density/fugacity, we have $\beta V_{F,CG}(r=0) \approx 1.955$ 
and $\beta V_{D,CG}(r=0) \approx 2.030$ for FD and DD potentials, 
respectively: repulsion increases by
11\% and 16\% in the two cases. Note that the observed differences between 
the DD and FD potentials are of the same order of the difference between
the zero-density potential and the state-dependent one, i.e. of the order
of the contribution of the many-body interactions, whose inclusion is the 
main motivation for considering state-dependent potentials.

The HNC procedure gives us estimates of $\Phi$ and of the virial pressure. 
To check these results, we perform grand canonical simulations of the 
CG model.  We measure the density as $\langle N \rangle/V$ 
and the pressure by using the standard virial expression. The results 
reported in Table~\ref{tableFD} are in good agreement with those (not
reported) obtained by
using the HNC approximation, confirming the good accuracy of the procedure.

\begin{table}[t]
\caption{Compressibility factor $Z = \beta P/\rho$ 
and polymer volume fraction $\Phi = 4 \pi \rho \hat{R}^3_g/3$ for a polymer 
CG system at two values of the fugacity. The reference polymer $Z(z)$ and 
$\Phi(z)$ have been obtain by using the equation of state reported 
in Ref.~\protect\onlinecite{Pelissetto-08}.
}
\label{tableFD}
\begin{tabular}{ccccccc}
\hline\hline
$z\hat{R}_g^3$ & $Z(z)$ & $Z_{CG}(z,0)$ & $Z_{CG}(z,z)$ &
     $\Phi(z)$ & $\Phi_{CG}(z,0)$ & 
     $\Phi_{CG}(z,z) $ \\
\hline
2.47 & 2.35 & 2.2920(1) & 2.3338(1) & 0.80 & 0.84104(4) & 0.77367(4) \\
77.7 & 3.90 & 3.7132(2) & 3.7819(2) & 1.50 & 1.68500(4) & 1.47603(4) \\
\hline\hline
\end{tabular}
\end{table}

Let us now compare the results with those of the original, full monomer
model.  As for the compressibility factor, the FD
model gives essentially the correct result for the lowest value of $z$
(the difference is less than 1\%), 
while it slightly underestimates (by 3\%) $Z$ for the largest value of $z$. 
In both cases, results obtained by using FD potentials are more 
accurate than those obtained by using the corresponding zero-density quantity.
The same conclusions are reached for $\Phi$, especially for the largest value
of $z$.
In the analysis of the low-density behavior, we observed that 
FD potentials are worse than zero-density potentials for what concerns the 
pressure.  Hence, we also consider
\begin{equation}
\hat{P} = \beta P \hat{R}_g^3 = {3\over 4\pi} Z \Phi .
\end{equation}
At the smallest value of $z$ we obtain $\hat{P} \approx 0.46020(3)$ by using 
the zero-density potential and $\hat{P} = 0.43105(3)$
by using the FD
potential, to be compared with the full-monomer result $\hat{P} \approx 0.45$.
The result obtained by using the zero-density potential 
is closer to the exact, full-monomer result, as expected on the basis of 
the low-density analysis. At
$z \hat{R}_g^3 = 77.7$ we have instead $\hat{P} \approx 1.4937(1)$ 
(zero-density), $\hat{P} \approx 1.3327(1)$ (FD), 
to be compared with the 
full-monomer results $\hat{P} \approx 1.40$. The state-dependent potential
provides now a result which is slightly more accurate.

From these results, it is clear that state-dependent potentials provide 
only approximations to pressure and density 
that worsen as the fugacity increases.
Accurate results can only be obtained by using 
CG multiblob models. If we use the tetramer model of 
Ref.~\onlinecite{DPP-12-b},
for $z\hat{R}_g^3=2.47$ we have $Z = 2.3457(5)$ and 
$\Phi = 0.7978(1)$, while for $z\hat{R}_g^3=77.7$ we have 
$Z = 3.8450(7)$ and $\Phi = 1.5257(1)$. Good agreement is observed for the 
lowest value of the fugacity, while small discrepancies occur for the
second one. If we use the decamer model\cite{DPP-12-b} we would obtain 
for $z\hat{R}_g^3=77.7$ the estimates 
$Z = 3.925(1)$ and $\Phi = 1.4984(3)$, which are now fully consistent with
the full monomer results.

\section{Conclusions} \label{sec6}

In this paper we have considered structurally derived 
state-dependent potentials for CG models,
with the purpose of understanding how useful they are in predicting
the thermodynamics of complex systems. As already discussed in 
Ref.~\onlinecite{Louis-02}, the virial or the Ascarelli-Harrison 
\cite{AH-69} pressure computed in the CG model
do not reproduce the pressure of the underlying system. 
But the same conclusions apply to essentially
all thermodynamic quantities. In the canonical ensemble Widom's
method does not give the correct underlying-model chemical potential, while in 
the grand canonical ensemble the virial pressure or $\langle N\rangle/V$
do not reproduce pressure and density of the underlying system. 
It is clear from the 
discussion that these results are not specific to these quantities. 
If the pair distribution function is used to define the state-dependent 
potentials, only the quantities that are directly related to the 
pair distribution function, like $K(\rho)$ or $H(z)$,  are identical 
in the CG and in the underlying model. All others differ. 
From a practical point of view, since state-dependent interactions are 
commonly used in large-scale simulations,
it is important to quantify the differences
between the CG estimates of the thermodynamic quantities 
and the underlying model values. In all cases we find that the passive 
approach provides the most accurate results. This is quite pleasant 
since the active-approach definitions
look quite unnatural: the pressure defined in the 
canonical ensemble does not satisfy Clausius' virial theorem---hence the 
pressure is not directly related to the average force on the boundaries---while
the density in the grand canonical ensemble defined by differentiating the 
grand potential differs from $\langle N\rangle/V$.

Ref.~\onlinecite{Louis-02} noted that the pressure of the underlying model 
can be obtained 
by using the {\em compressibility route}. However, it should 
be clear that one is referring to the compressibility route 
in the underlying model, i.e., to Eq.~(\ref{pFA-KFA}).
But this means that one
is obtaining the pressure from a quantity computed in the underlying
microscopic model,
hence the DD potential plays no role. Analogously, 
in the grand-canonical ensemble, the correct density can be obtained 
by integrating $H(z)$, but again this calculation 
does not really make use of the model with FD potentials.

It is important to stress that structurally derived 
state-dependent potentials depend on the 
ensemble one uses, or rather on the thermodynamic variable one uses
to identify the thermodynamic state of the system. At a given 
thermodynamic state of the microscopic system, DD potentials and 
FD potentials differ, since they correspond to 
CG monoatomic systems that have the same pair distribution function
but different densities and chemical potentials.
Therefore, DD canonical potentials should not be used 
to predict phase lines, since the coexistence analysis is typically performed 
in other ensembles.

Although in this paper we have only discussed DD and FD
potentials, the results are general and apply to other ensembles, 
for instance to pressure
dependent potentials in the isothermal-isobaric ensemble. 
However, the conclusions
do not necessarily apply to potentials that are only temperature dependent. 
If the thermodynamic state one is considering 
can be connected by a thermodynamic 
path at fixed temperature to the zero-density state (this is always
true in the absence of phase transitions or at least in the low-density
phase), since temperature does not appear explicitly in the integrations 
leading to the free energies,
the free energy is essentially correct (one should only add the zero-density 
contribution of a single, isolated molecule). 
Hence, all thermodynamic quantities that 
can be obtained from the free energies are correct. 
An interesting method for transferring CG potentials between temperatures 
in the context of the multiscale coarse-graining
force-matching method is described in 
Ref.~\onlinecite{KNV-09}.
It is important to note that the ensemble dependence does not occur 
in CG models defined by using the multiscale coarse-graining approach based on 
the force-matching method.\cite{force-matching1}
Indeed, the distribution of the atomistic configurations is obviously
ensemble independent in the infinite-volume limit, hence also 
the matching conditions and the CG model do not depend on the 
ensemble (see Ref.~\onlinecite{DA-10} for a discussion in the 
isobaric-isothermal ensemble).

Our results also shed some light on mixed approaches in which one matches 
the pair distribution and some other thermodynamic property. 
These approaches are conceptually correct only if the thermodynamic 
quantity is appropriately chosen. In the canonical ensemble, the only
correct choice is the isothermal compressibility, i.e. the quantity $K$
we have introduced, while in the grand canonical ensemble one should use
$H(z)$. It is also possible to enforce the value of the pressure, i.e.
to work in the isothermal-isobaric ensemble, but in this case one should relax 
the condition that the microscopic system and the CG model have the 
same density. For each choice, a different state-dependent potential 
is obtained, as observed, for instance, in Ref.~\onlinecite{WJK-09} 
in the context of CG models of water. Note that these observations
can be trivially generalized to multicomponent systems. 
In this case, one can analogously enforce the Kirkwood-Buff\cite{KB-51}
equations that relate thermodynamic properties to integrals of 
distribution functions, see Ref.~\onlinecite{GMJV-12} for a recent example.

The fact that DD potentials are unable to provide
the correct results for the thermodynamic quantities should not 
be surprising. After all, even with the use of 
a state-dependent potential, the CG model is still an approximation to the 
original microscopic model. Hence, one should consider this approach as a 
simple method to obtain relatively good approximations for the 
behavior of complex systems. 

It is easy to find situations, both 
in theoretical or experimental work, in which 
state-dependent potentials are useful. For instance, one could consider 
a system of complex molecules for which simulations are particularly difficult.
In order to obtain with reasonable precision the value of the 
thermodynamic quantities in a range of densities, one could 
perform a simulation 
at a density $\rho_p$, measuring all thermodynamic quantities
and the density-dependent potential $V_{D,CG}({\bf r};\rho_p)$. Then, 
to compute the pressure for $\rho\not=\rho_p$,
one could write 
\begin{eqnarray}
   P(\rho) = P(\rho_p) + [P(\rho) - P(\rho_p)],
\end{eqnarray}
and determine the second term by using the CG model that 
uses the DD potential computed at density $\rho_p$. Hence, one could
estimate 
\begin{eqnarray}
   P(\rho) = P(\rho_p) + [P_{CG}(\rho,\rho_p) - P_{CG}(\rho_p,\rho_p)],
\label{approx}
\end{eqnarray}
using the CG model to compute the differences between the state point of 
interest and that for which exact results, i.e. results obtained in the 
microscopic model, are available. 
CG models with zero-density potentials use exactly this strategy, 
fixing $\rho_p = 0$. The same strategy could also be used in experiments,
allowing one to obtain all thermodynamic informations in a range of 
densities, using experimentally determined thermodynamic and structural 
data at $\rho_p$.

We have checked this strategy in our polymer example. Assuming that the 
exact results for $P$ and $\mu$ are available at $\Phi = 0.8$,
we wish to compute $P$ and $\mu$ at $\Phi = 1.5$. Using the
DD potential computed at $\Phi = 0.8$, we obtain 
$Z = 2.44790(4)$ and $\beta \mu^{\rm exc} = 2.8108(3)$ 
at $\Phi = 0.8$ (see Table~\ref{tableDD}) and
$Z = 3.8332(4)$ and $\beta \mu^{\rm exc} = 5.4962(4)$ at $\Phi = 1.5$.
Hence, we would estimate at $\Phi = 1.5$
\begin{eqnarray}
Z =  3.74  \qquad\qquad
\beta \mu^{\rm exc} =  5.25,
\label{est-phen}
\end{eqnarray}
to be compared with the FM predictions 
$Z \approx 3.90$ and $\beta \mu^{\rm exc}  = 5.38$. 
Estimates (\ref{est-phen}) are significantly more precise than those 
reported in Table~\ref{tableDD}, confirming the usefulness of the approach.

Finally, note that we have considered here very simple CG models 
in which the CG molecules are monoatomic. In practical soft-matter 
applications CG models are usually polyatomic.
It is quite obvious that the 
same conclusions apply to these more complex models: also in this case
structurally derived potentials depend on the ensemble and thermodynamics is 
only approximately reproduced.

\section*{Acknowledgments}

The authors thank Barbara Capone and Luigi Delle Site for useful comments.
C.P. is supported by the Italian Institute of Technology (IIT) under the
SEED project grant number 259 SIMBEDD – Advanced Computational Methods for
Biophysics, Drug Design and Energy Research.


\begin{thebibliography}{99}
\bibitem{fixpress}
In the full-monomer model we have $\beta P \hat{R}^3_g = 0.45$ and 
$\beta {P} \hat{R}^3_g = 1.40$ for $\Phi = 0.80$ and $1.50$. 
For comparison, we report 
the pressure-dependent potentials for these two values of the pressure.
Using the HNC approximation we can also compute the thermodynamic 
variables using these state-dependent interactions. We obtain 
$\Phi = 0.784$, $Z = 2.397$, $\beta \mu^{\rm exc} = 2.72$ for 
$\beta {P} \hat{R}^3_g = 0.45$ and 
$\Phi = 1.484$, $Z = 3.942$, $\beta \mu^{\rm exc} = 5.71$ for 
$\beta {P} \hat{R}^3_g = 1.40$.
\end{thebibliography}
\end{document}